# Designing for Collaborative Sensemaking: Leveraging Human Cognition For Complex Tasks

Participant: Nitesh Goyal, Supervisor: Susan R. Fussell
Cornell University
Dept. of Information Science
Ithaca, NY 14850, USA
`ngoyal@cs.cornell.edu, sfussell@cornell.edu`

**Research Area.** Collaborative Sensemaking, Crowdsourcing, User Interfaces for Web Applications

**Research Topic**. My research aims to design systems for complex sensemaking by remotely located non-expert collaborators (crowds), to solve computationally hard problems like crimes.

## 1  Research Problem and Related Work

While sensemaking has been studied in the past, designing interfaces for relatively complex sensemaking where experts and non-experts may collaborate remains a challenge. I am interested in how such collaborative sensemaking by crowds (non-experts) may help us better solve unstructured problems, where traditional computational techniques have failed. In particular, I am researching how to design for experts and non-experts (crowds) in the crime-solving domain, where solutions are often found through serendipity, instead of rules.

Crowdsourcing for somewhat complex tasks has been pursued in the past. Collaborative document editing in Soylent [2], creating taxonomy of colours in Cascade [3], suggesting a travel itinerary under constraints using Mobi [4], and mining sentiments by crowds for text analytics in OpinionBlocks [12] are some recent forays where crowdsourcing has shown to be performant and/or efficient. However, more open-ended domains like crime solving, requiring serendipitous discovery of clues and criminals, have yet to be crowd-sourced successfully.

As number of workers and associated workflows grow in complexity, crowdsourcing can be challenging [1]. Crowdsourcing for complex workflows has been pursued also. For example, CrowdForge by Kittur et al [5] explains how map-reduce framework popularized by Google may be used to partition bigger complex tasks into smaller tasks dynamically by workers. Further, Malone et al 's [6] aggregation dimension suggests that crowd-workers can either work alone independently or depend upon each other to work together. Little et al [7]'s

TurKit can further help decide what to present to each worker such that the flow of results of tasks between dependent workers can be controlled.

As such crowd-workflows become complex, researchers must identify the level of crowd-supervision needed for optimal output. Kulkarni et al [8] designed Turkomatic based on price-divide-loop such that real time visualization of the workflow-design is evident because unsupervised crowds failed to produce proper workflows resulting in a less than optimal output. On the other hand, supervised crowds in a conversational-agent, Chorus by Lasecki et al [9] made users believe that a single user exist behind Chorus. Instead, Chorus employs multiple crowd workers who collectively create response possibilities, such that Crowd workers can learn and remember collectively.

To summarize, while relatively complex tasks and workflows using crowds have been attempted, we have yet been unable to design a system that may structure non-experts (lesser trained crowds) and experts (trained workers) together in an interface to solve complex challenges like crime solving.

## 2   Research Hypothesis

The core aim of this research is to pursue a user centred design approach to designing a web-interface, and underlying system, that may enable collaborations between experts and non-experts, and within non-experts. I hypothesize that a user-centered-designed web-interface that supports collaboration between non-expert crowds for solving carefully broken down micro-tasks would help leverage distributed human crowd-cognition to solve complex tasks like crimes.

## 3   Proposed Method

I plan to integrate my findings based on a mixed-methods study. First, I will understand how trained-non-experts (trained students, through video and usage-log analysis) solve complex problems singularly and collaboratively. Consequently, I will extract important features that result in success and failure in problem solving. Based on these features, I propose to create a web-interface for collaborative problem solving. Finally, I will design a study to validate whether the identified features lead to success or failure with non-expert crowds? So, based on the iterative nature of design process, my proposed solution would involve multiple iterations and steps before I design the final interface:
Step 1. Understand role of currently used features for solo sensemaking.
Step 2. Explore effects of information-sharing collaborative sensemaking.
Step 3. Extract features to identify micro-tasks, and workflows for success.
Step 4. Design web-interface for experts and non-experts to collaborate.
Step 5. Design a set of user-studies to measure user-experience, and performance achieved by non-experts with the web-interface at solving crimes.

## 4   Progress So-far

I have completed two iterations of system building of SAVANT tool to support solo [10] and collaborative sensemaking [11, 12] to better understand role of different design features:

In Iteration 1 (Step 1), I tested the utility of system-generated visualization of data links and a notepad for collecting annotations, and found system-generated visualizations to be significantly important in solving crimes [10].

In Iteration 2 (Step 2), I explored value of implicitly sharing insights by self-created visualizations of annotations, without explicitly pushed/requested information by collaborators. When implicit sharing of notes and self-created visualization of these notes was available, users identified more clues [11, 12].

## 5   Next Steps

I am presently working on Task Analysis for Step 3. I am conducting a qualitative video-analysis, and usage-log analysis of the actions performed by successful and unsuccessful pairs in Step 2. Based on video-analysis, 3 *design goals* seem promising for success: *externalizing insights; shoe-boxing visually; and iterating over previously collected information*. I am also identifying user-actions, based on interface-log analysis, when pursued multiple times by users would lead to successful resolution of the task.

Based on preliminary findings, I am designing the web-interface (SAVANT) for non-experts and experts with recommended steps associated with success. Based on these findings, I'd be better equipped with knowledge of micro-tasks, workflows, and atomicity of the dataset that would enable success in task-resolution. So, I'd propose using the lessons learnt to design the next SAVANT version where users using the full SAVANT suite might be able to collaborate and auto-direct micro-tasks to crowds that would support/challenge their own insights and help resolve the crime-solving task.

Finally, I would like to use the INTERACT 2015 Doctoral Consortium to help design user-studies and determine measures to evaluate user-experience, and features of SAVANT like system-generated micro-tasks vs. manually generated micro-tasks; and presence of messaging with the crowds vs. no messaging.

## 6   Proposed Solution

My proposed solution would include features identified in Section 4, and more importantly would leverage the interaction between these features to aid collaborative sensemaking. These features are: *visualizations: system generated, and user-generated; and information sharing: implicitly by the system, and user-mediated*. Further, based on the *design goals* identified in Section 5, the proposed solution would suggest recommendations to non-experts to aid sensemaking.

## 7  Expected Contributions

This doctoral work will have design contribution of creating an interface where micro-tasks may be auto-generated for the crowds and results produced are fed back into the original system for analysis. Theoretical contribution of my work would be understanding how micro-task based crowd-supported system may enhance task-resolution of complex tasks, and support collaboration.